\documentclass[twoside,twocolumn,9pt]{article}
\usepackage{extsizes}
\usepackage[super,sort&compress,comma]{natbib} 
\usepackage[version=3]{mhchem}
\usepackage[left=1.5cm, right=1.5cm, top=1.785cm, bottom=2.0cm]{geometry}
\usepackage{balance}
\usepackage{times,mathptmx}
\usepackage{sectsty}
\usepackage{graphicx} 
\usepackage{lastpage}
\usepackage[format=plain,justification=justified,singlelinecheck=false,font={stretch=1.125,small,sf},labelfont=bf,labelsep=space]{caption}
\usepackage{float}
\usepackage{fancyhdr}
\usepackage{fnpos}
\usepackage[english]{babel}
\usepackage{array}
\usepackage{droidsans}
\usepackage{charter}
\usepackage[T1]{fontenc}
\usepackage[usenames,dvipsnames]{xcolor}
\usepackage{setspace}
\usepackage[compact]{titlesec}

\usepackage{epstopdf}

\definecolor{cream}{RGB}{222,217,201}

\begin{document}

\pagestyle{fancy}
\thispagestyle{plain}
\fancypagestyle{plain}{

\renewcommand{\headrulewidth}{0pt}
}

\makeFNbottom
\makeatletter
\renewcommand\LARGE{\@setfontsize\LARGE{15pt}{17}}
\renewcommand\Large{\@setfontsize\Large{12pt}{14}}
\renewcommand\large{\@setfontsize\large{10pt}{12}}
\renewcommand\footnotesize{\@setfontsize\footnotesize{7pt}{10}}
\makeatother

\renewcommand{\thefootnote}{\fnsymbol{footnote}}
\renewcommand\footnoterule{\vspace*{1pt}%
\color{cream}\hrule width 3.5in height 0.4pt \color{black}\vspace*{5pt}} 
\setcounter{secnumdepth}{5}

\makeatletter 
\renewcommand\@biblabel[1]{#1}            
\renewcommand\@makefntext[1]%
{\noindent\makebox[0pt][r]{\@thefnmark\,}#1}
\makeatother 
\renewcommand{\figurename}{\small{Fig.}~}
\sectionfont{\sffamily\Large}
\subsectionfont{\normalsize}
\subsubsectionfont{\bf}
\setstretch{1.125} 
\setlength{\skip\footins}{0.8cm}
\setlength{\footnotesep}{0.25cm}
\setlength{\jot}{10pt}
\titlespacing*{\section}{0pt}{4pt}{4pt}
\titlespacing*{\subsection}{0pt}{15pt}{1pt}

\fancyfoot{}
\fancyfoot[RO]{\footnotesize{\sffamily{1--\pageref{LastPage} ~\textbar  \hspace{2pt}\thepage}}}
\fancyfoot[LE]{\footnotesize{\sffamily{\thepage~\textbar\hspace{3.45cm} 1--\pageref{LastPage}}}}
\fancyhead{}
\renewcommand{\headrulewidth}{0pt} 
\renewcommand{\footrulewidth}{0pt}
\setlength{\arrayrulewidth}{1pt}
\setlength{\columnsep}{6.5mm}
\setlength\bibsep{1pt}

\makeatletter 
\newlength{\figrulesep} 
\setlength{\figrulesep}{0.5\textfloatsep} 

\newcommand{\topfigrule}{\vspace*{-1pt}%
\noindent{\color{cream}\rule[-\figrulesep]{\columnwidth}{1.5pt}} }

\newcommand{\botfigrule}{\vspace*{-2pt}%
\noindent{\color{cream}\rule[\figrulesep]{\columnwidth}{1.5pt}} }

\newcommand{\dblfigrule}{\vspace*{-1pt}%
\noindent{\color{cream}\rule[-\figrulesep]{\textwidth}{1.5pt}} }

\makeatother

\twocolumn[
  \begin{@twocolumnfalse}
\vspace{3cm}
\sffamily
\begin{tabular}{m{4cm} p{14cm} }

 & \noindent\LARGE{\textbf{Stiffness of the C-terminal disordered linker affects the geometry of the active site in endoglucanase Cel8A }} \\
\vspace{0.3cm} & \vspace{0.3cm} \\

 & \noindent\large{Bartosz R\'{o}\.{z}ycki,$^{\ast}$\textit{$^{a}$} and Marek Cieplak\textit{$^{a}$}} \\

 & \noindent\normalsize{Cellulosomes are complex multi-enzyme machineries which efficiently degrade plant cell-wall polysaccharides. The multiple domains of the cellulosome proteins are often tethered together by intrinsically disordered regions. The properties and functions of these disordered linkers are not well understood. In this work, we study endoglucanase Cel8A, which is a relevant enzymatic component of the cellulosomes of {\sl Clostridium thermocellum}. We use both all-atom and coarse-grained simulations to investigate how the equilibrium conformations of the catalytic domain of Cel8A are affected by the disordered linker at its C terminus. We find that when the endoglucanase is bound to its substrate, the effective stiffness of the linker can influence the distances between groups of amino-acid residues throughout the entire enzymatic domain. In particular, variations in the linker stiffness can lead to small changes in the geometry of the active-site cleft. We suggest that such geometrical changes may, in turn, have an effect on the catalytic activity of the enzyme. } \\

\end{tabular}

  \end{@twocolumnfalse} \vspace{0.6cm}

  ]

\renewcommand*\rmdefault{bch}\normalfont\upshape
\rmfamily
\section*{}
\vspace{-1cm}


\footnotetext{\textit{$^{a}$~Institute of Physics, Polish Academy of Sciences, Al. Lotnik{\'o}w 32/46, 02-668 Warsaw, Poland}}
\footnotetext{$^{\ast}$~Tel: 48 22 116 3265; E-mail: rozycki@ifpan.edu.pl}

\footnotetext{\dag~Electronic Supplementary Information (ESI) available: []. See DOI: 10.1039/b000000x/}




\section{Introduction}

Cellulosomes are large, multi-enzyme complexes that are found on the surfaces of different 
cellulolytic microorganisms \cite{Bayer1999,Bayer2004}. 
Their function is degradation of plant cell-wall polysaccharides 
to simple sugars \cite{Bayer2004,Bayer2007}. 
One of the major challenges in the current research on the cellulosomes is to understand 
how their molecular architectures and activities are related \cite{Bayer2007,Bayer2013}. 
Importantly, the structural characterization of the cellulosome complexes can 
provide a platform for biotechnological and nanotechnological applications, 
including prospects for producing biofuels from plant-cell-wall biomass \cite{Bayer2007}. 

The multiple proteins of cellulosomes are composed of numerous functional domains, 
which interact with each other and with polysaccharides. 
These domains are usually tethered to one another by intrinsically disordered 
polypeptide segments, which are commonly called linkers. 
Interestingly, a number of distinct multi-domain proteins and protein complexes 
belonging to cellulosomes of different cellulolytic microorganisms 
have been found to be highly flexible in solution 
\cite{Hammel2005,Czjzek_review2012,Currie2012,JBC2013}. 
Their flexibility seems to be needed for the simultaneous binding of the multiple 
enzymatic domains to plant cell-wall polysaccharides. 
A likely role of the disordered linkers is thus to provide the required 
degree of flexibility while maintaining the integrity of the cellulosome complexes.

The intrinsically disordered linkers in cellulosome complexes differ significantly in amino-acid 
sequence and length -- from several to about a hundred amino-acid residues. 
The reasons for this variability among the cellulosomal linkers have been a puzzle so far. 
As a matter of fact, the biological functions of the disordered linkers are 
not well understood. It has been demonstrated that the sequential length of linkers 
can influence the enzymatic activity of designer cellulosomes \cite{Bayer2013} 
but in some cases this effect is marginal \cite{Caspi2009}. 
However, the relations between the physical properties of the disordered linkers and 
the enzymatic activity of the cellulosomes remain unknown. 
Here, we study endoglucanase Cel8A of {\sl Clostridium thermocellum} and show that 
effective stiffness of the disordered linker at the C-terminus of the catalytic 
domain has an impact on the geometry of the active-site region when 
Cel8A is bound to its substrate. Such geometrical changes may in turn 
influence the catalytic activity of the enzyme. 

A kinetic model for a two-domain cellulase, 
which treats the catalytic domain (CD) and the carbohydrate-binding module (CBM) 
as two random walkers in one dimension, 
predicts that the cooperative action of the CD and CBM domains depends on 
the length and stiffness of the linker between CD and CBM \cite{Makarov2009}. 
This model is quite general and does not make use of 
structural information about the cellulase. 
In contrast, our study is based on the atomic structure of the catalytic domain of 
endoglucanase Cel8A, and gives predictions about the structural rearrangements within the enzyme. 
We use coarse-grained models in which amino-acid residues 
are represented as single beads, and investigate how the near-native conformations of 
the CD in Cel8A are affected by the stiffness of the disordered linker at its C terminus. 
The predictions of the coarse-grained model are corroborated by all-atom simulations. 

\begin{figure}[t]
\begin{center}
\scalebox{0.4}{\includegraphics{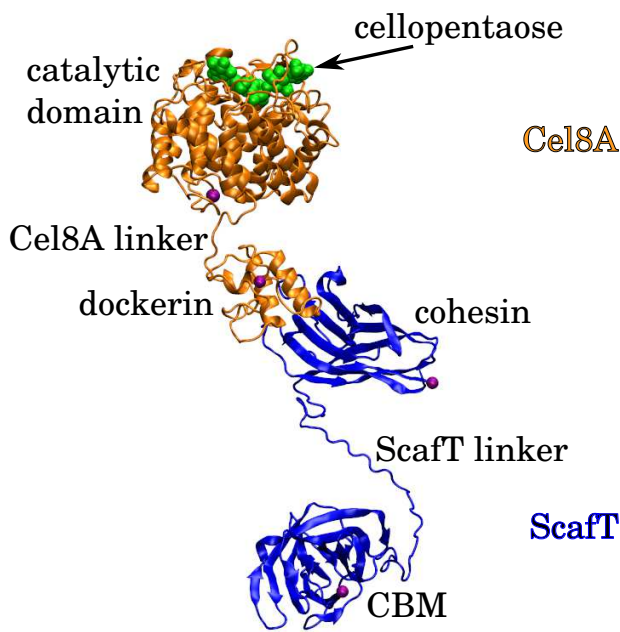}}
\caption{\label{complex_struc} 
Structure of the full-length Cel8A-ScafT complex. Cel8A and ScafT proteins are shown 
in orange and blue, respectively. The substrate molecule (cellopentaose) is shown in green. 
The terminal residues of Cel8A and ScafT are marked as purple spheres. 
The functional domains and disordered linkers are indicated. 
The sequential lengths of the Cel8A and ScafT linkers are 17 and 25, respectively.
}
\end{center}
\end{figure}

The use of thermostable cellulosomes is advantageous for the breakdown of
cellulosic biomass toward the commercial production of biofuels. For this reason, 
experimental and bioinformatics-based approaches have been combined to engineer mutants of
endoglucanase Cel8A that are thermally more stable and catalytically more active at
elevated temperatures than the wild-type Cel8A \cite{AnbarBayer2010,AnbarBayer2012}. 
In particular, a quadruple mutant has been shown to exhibit a 14-fold increase in 
the half-life of activity at 85$^{\circ}$C \cite{AnbarBayer2012}. This mutant, herein 
referred to as Cel8A*, contains the following four mutations within its CD 
as compared to the wild-type protein: K276R, G283P, S329G, S375T. 
Interestingly, despite the enhanced activity and thermostability of Cel8A* as a free enzyme, 
its substitution for the wild-type endoglucanase Cel8A within the cellulosome context 
deteriorates overall degradation of cellulose \cite{Bayer_CR_2014}. 
For example, the wild-type endoglucanase Cel8A in complex with scaffoldin ScafT 
has been demonstrated to exhibit a higher enzymatic activity than 
the Cel8A*-ScafT designer cellulosome \cite{Bayer_CR_2014}. 
These discoveries show that physical interactions between different cellulosomal units 
can affect the overall enzymatic activity of cellulosomes in a non-trivial way. 
Here, we investigate what kinds of inter-domain interactions can affect the structure and 
activity of the Cel8A-ScafT mini-cellulosome, which is depicted in Fig.~\ref{complex_struc}. 
Our results indicate that linker-mediated interactions can influence 
the structure of the active site in Cel8A. 
We identify differences in the geometry of the substrate-binding region between Cel8A and Cel8A*. 
We also show that the effective stiffness of the C-terminal linker influences 
the structure of this region in Cel8A* in a way which is different than in Cel8A.

\section{Results}

Endoglucanase Cel8A is a glycoside hydrolase of family 8. 
We consider the endoglucanase Cel8A from {\sl C. thermocellum}, which 
comprises its CD and a type I dockerin at the C-terminus. 
Cel8A is shown in orange in Fig.~\ref{complex_struc}. 
The CD binds to cellulose chains and cleaves their $(1 \to 4)$-$\beta$-D-glucosidic linkages. 
The function of the dockerin domain is to bind tightly to a complementary 
cohesin domain and, thus, anchor Cel8A to a cellulosomal scaffoldin subunit. 
The linker between the CD and the dockerin is a proline-rich peptide segment. 
ScafT is a section of the {\sl C. thermocellum} cellulosomal CipA scaffoldin subunit \cite{Bayer_CR_2014}. 
It is shown in blue in Fig.~\ref{complex_struc}. 
ScafT contains the family 3a CBM and a single cohesin domain of type I, namely, cohesin 3. 
The linker between the CBM and the cohesin contains multiple proline and threonine residues. 
The cohesin binds tightly and specifically to the Cel8A dockerin domain. 
The function of the CBM is to bind to cellulose chains and, thus, bring 
the cellulosome to the vicinity of its substrate. 

In subsection~\ref{Inter_domain_Cel8A-ScafT} we consider 
the Cel8A-ScafT complex in a free state, {\sl i.e.}, 
when neither the CD nor the CBM are bound to cellulose chains. 
We next consider a bound state, in which both the CD and the CBM are anchored to 
cellulose chains. Interestingly, the two 'anchors' impose restraints on 
the motions of the Cel8A-ScafT complex. 
In subsections~\ref{intra_domain_Cel8A} and \ref{intra_domain_Cel8A*} we study 
the effect of these restraints on the conformations of 
the CD in Cel8A and Cel8A*, respectively. 
We find that an important factor is the effective stiffness of the linker 
between the CD and the remaining parts of the mini-cellulosome in the bound stare. 
In subsection~\ref{linker_stiffness} we estimate the degree of stiffness of 
cellulosomal linkers.

\subsection{Inter-domain contacts in Cel8A-ScafT \label{Inter_domain_Cel8A-ScafT} }

To examine the physical interactions between the different domains within 
the free Cel8A-ScafT complex, we use a coarse-grained model 
introduced by Kim and Hummer \cite{Hummer2008}. 
In this model, folded protein domains are treated as rigid bodies and 
the disordered linkers are simulated as chains of amino-acid beads. 
This model has been successfully applied to various multi-domain enzymes, 
including xylanases \cite{XynZ2015}, protein kinases \cite{pkc2011} and 
kinases in dynamic complexes with phosphatases \cite{ERK2HePTP_2011,p38_2011}. 
It involves statistical contact potentials and 
its description is given in Methods section \ref{methods_CG}. 
Within the framework of this coarse-grained model, we performed Replica Exchange 
Monte Carlo simulations of both the wild-type Cel8A-ScafT complex and 
the quadruple mutant Cel8A*-ScafT mini-cellulosome. 
In these simulations we assumed that the four mutations 
had only a local effect on the structure of the catalytic domain of Cel8A*. 
Specifically, our assumption here is that the structures of the catalytic domains of 
Cel8A and Cel8A* differ only in the chemical composition and orientations of the side chains 
-- and, thus, in the energy of the inter-domain contacts -- 
while the structures of the backbone chains of Cel8A and Cel8A* are identical. 
Such assumptions are often made in theoretical and computational studies on 
the effects of mutations on protein stability \cite{foldX} 
but do not have to be correct in general. 

The simulation results are presented in Supplementary Information. 
They show that the distributions of the maximum extension for 
the wild-type Cel8A-ScafT complex and the quadruple mutant Cel8A*-ScafT 
mini-cellulosome are practically identical (see Fig.~S1A). 
Also, there are practically no differences in the distributions of 
the radius of gyration between Cel8A-ScafT and Cel8A*-ScafT (see Fig.~S1B). 
Moreover, we do not find any systematic differences neither in 
the probabilities of the inter-domain contacts (see Tab.~SI) nor in 
the end-to-end distance distributions for the flexible linkers (see Fig.~S1C and D).  
In summary, the conformational ensembles of Cel8A-ScafT and Cel8A*-ScafT are 
found to be practically indistinguishable in these simulations. 

Out of the four mutated residues in Cel8A*, two are embedded in 
the helical core of the CD (G283P and S375T) and 
two are in loops on the protein surface (K276R and S329G),  
as indicated in Fig.~\ref{contacts_pic}. 
Moreover, two of the four mutations (K276R and S375T) do not alter 
the general physicochemical properties of the original residues. 
Therefore, it is not unreasonable that the four mutations in the catalytic domain 
do not affect the inter-domain contacts within the Cel8A-ScafT complex. 
On the other hand, it has been demonstrated that Cel8A-ScafT is more active than 
Cel8A*-ScafT despite the fact that Cel8A* is more active that Cel8A \cite{Bayer_CR_2014}. 
Therefore, the interactions between Cel8A and ScafT are relevant for 
the catalytic activity of the enzyme. 
These interactions can be direct inter-domain contacts, 
as investigated here within the framework of the Kim-Hummer model, 
or indirect linker-mediated interactions.

\subsection{Intra-domain contacts within the catalytic domain of Cel8A \label{intra_domain_Cel8A} }

\subsubsection{Results of coarse-grained structure-based simulations}

The Kim-Hummer model does not take into account any structural rearrangements of 
the individual domains which may be caused by direct domain-domain interactions
or by linker-mediated interactions within the protein complex. 
To study the latter effect, we use a different coarse-grained model, 
which is structure-based \cite{Hoang2,SikoraPLOS2009,Proteins2014,Poma2015} 
and which is described in detail in Methods section \ref{methods_Go}. 

\begin{figure*}[t]
\begin{center}
\scalebox{0.4}{\includegraphics{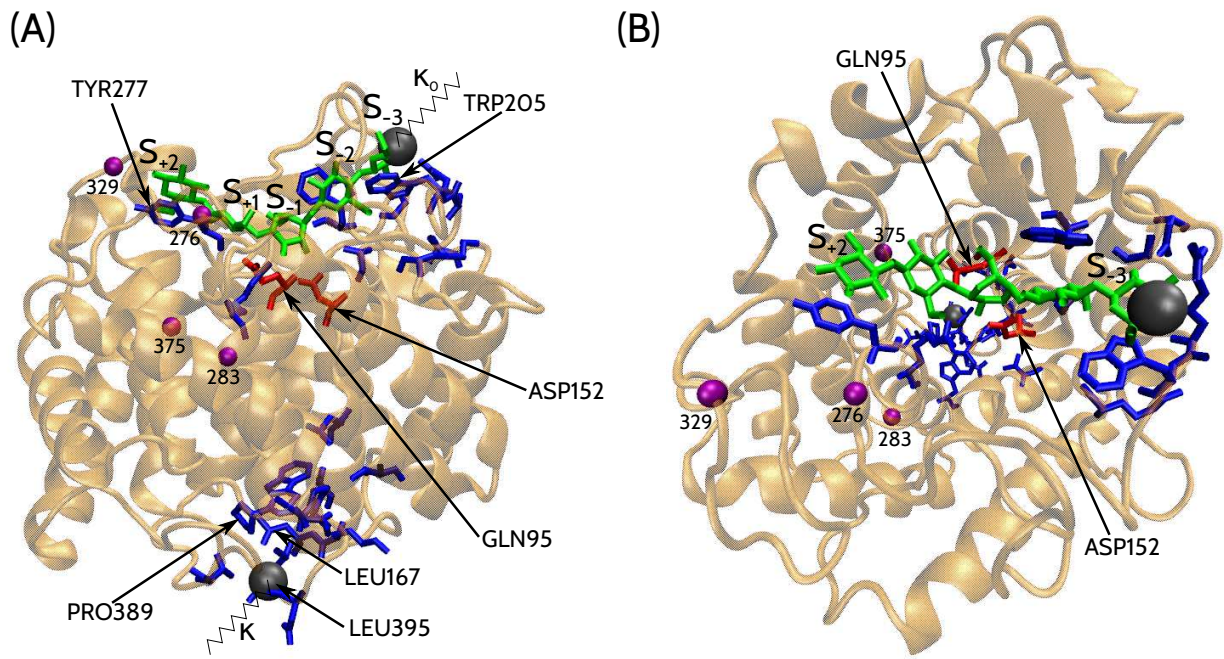}}
\caption{\label{contacts_pic} 
Structure of the catalytic domain in complex with its substrate. (A) Side view. 
The gray spheres mark the centers to which the virtual springs are attached: 
the first one to atom C4 of the glucose ring S$_{-3}$ (upper sphere) and 
the second one to the $\alpha$-C atom of LEU395 (lower sphere), which is 
the C-terminal residue of this domain structure. 
The corresponding spring constants are denoted by $\kappa_0$ and $\kappa$. 
The blue sticks mark the amino-acid residues forming the native contacts 
that are predicted by the coarse-grained simulations 
to shorten when the spring is applied to the C-terminal residue. 
The residues indicated by arrows correspond to the data shown in Fig.~\ref{dij}. 
The red sticks mark the catalytically active residues with the sequence numbers 95 and 152. 
The green sticks mark the substrate. The glucose units are labeled by S$_{-3}$, S$_{-2}$, 
S$_{-1}$, S$_{+1}$ and S$_{+2}$ as in Ref.~\cite{1KWF}. 
The chemical reaction catalyzed by the endoglucanase results in 
breaking the bond between S$_{-1}$ and S$_{+1}$. 
Unit S$_{-3}$ connects the substrate to a cellulose chain. 
The purple spheres mark the mutation sites with the sequence numbers 276, 283, 329 and 375 
as indicated by the labels. (B) Top view. The color code is the same as in panel A. } 
\end{center}
\end{figure*}

The atomic structure of the catalytic domain of Cel8A in complex with a cellopentaose molecule 
has been solved \cite{1KWF} and deposited in the Protein Data Bank (PDB code 1KWF). 
It is depicted in Fig.~\ref{contacts_pic}. 
Using the overlap criterion \cite{Wolek}, we determine both the amino-acid contacts and 
the contacts between the glucose units and the amino acids within this atomic structure. 
The overlap criterion predicts 1112 contacts between the amino-acid residues 
and 30 contacts between the enzyme and its substrate. 
In the framework of the structure-based coarse-grained model which we use in this study, 
both the amino-acid residues and the glucose units are represented as single beads 
that are centered at the $\alpha$-C and C4 atoms, respectively. 
All of the native contacts are described by the Lennard-Jones potential 
acting between the residue beads. Details are given in Methods section \ref{methods_Go}. 

The cellopentaose consists of five glucose units which are denoted by S$_{-3}$, 
S$_{-2}$, S$_{-1}$, S$_{+1}$ and S$_{+2}$ as indicated in Fig.~\ref{contacts_pic}. 
The cellopentaose-binding site in the CD of Cel8A is in the shape of a cleft. 
The catalytically active residues have the sequence numbers 95 and 152. 
They are located at the center of the active-site cleft and 
form contacts with the glucose units S$_{-1}$ and S$_{+1}$. 
Unit S$_{-3}$ connects the cellopentaose to a cellulose chain. 
To mimic this connection in our model, a virtual spring is attached to 
the bead representing the glucose unit S$_{-3}$, see Fig.~\ref{contacts_pic}. 
The Cel8A-ScafT complex binds to cellulose chains 
also at a different spot {\sl via} its CBM domain. 
To mimic the linker between the CD and the remaining part of 
the cellulose-bound Cel8A-ScafT complex, 
another virtual spring is used. Specifically, it is 
attached to the LEU395 residue bead at the C-terminus of the CD. 
These two springs restrain the thermal motion of the CD. 
The stiffness of the first spring (applied to unit S$_{-3}$) is set to  
$\kappa_0 = 160$~kcal/(mol~{\AA}$^{2}$), which is equal the stiffness of 
the pseudo-bond between consecutive beads in our Go-type model. 
This parameter, denoted by $\kappa_0$, is fixed in all of the simulations we report here. 
The stiffness of the second spring (applied to LEU395) is denoted by $\kappa$. 
It depends in general on the amino-acid sequence and length of the disordered linkers. 
It is a variable of our model. 

We performed Molecular Dynamics (MD) simulations of the system described above 
for several different values of the spring constant $\kappa$ between 0 and 8~kcal/(mol~{\AA}$^{2}$). 
This range of values should capture the persistence lengths of disordered linkers with 
diverse sequences, as we argue in section~\ref{linker_stiffness} below. 
In order to accommodate this wide range, we use a logarithmic scale for $\kappa$ in the figures. 

\begin{figure}[h]
\begin{center}
\scalebox{0.7}{\includegraphics{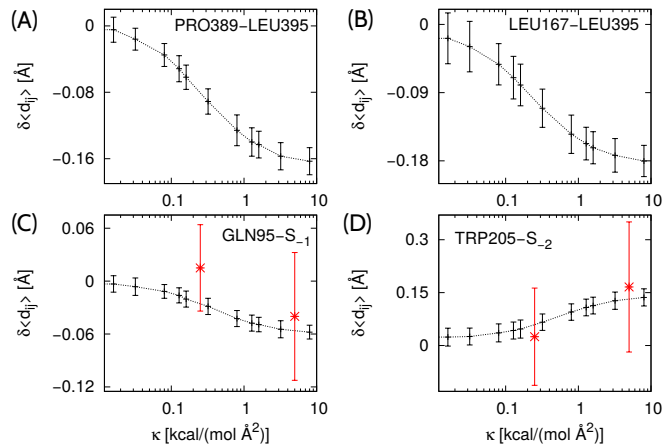}}
\caption{\label{dij} 
Variations in the average distance between residues forming native contacts, 
$\delta \langle d_{ij} \rangle = 
\langle d_{ij} \rangle ( \kappa = 0 ) - \langle d_{ij} \rangle ( \kappa > 0 )$, 
as a function of the stiffness $\kappa$ of the linker-imitating spring. 
The residue pairs forming the native contacts are indicated. 
These residues are indicated by arrows in Fig.~\ref{contacts_pic}. 
Panels A and B refer to two contacts with the C-terminal residue (LEU395). 
Panels C and D correspond to two contacts within the catalytic cleft. 
The error bars indicate variations between results from nine independent trajectories. 
The data points in red correspond to the results of the 30~ns all-atom MD simulations. 
Note that the logarithmic scale is used on the horizontal axes. }
\end{center}
\end{figure}

Interestingly, the stiffness of the spring attached to the C-terminus, $\kappa$, is observed 
to have a non-local effect on the average distances $\langle d_{ij} \rangle$ 
between the residues forming the native contacts. 
Examples are shown in Fig.~\ref{dij}. 
Panel A shows how the average distance between PRO389 and LEU395 varies with $\kappa$. 
The latter residue is at the C-terminus, where the second spring is applied. 
Panel B refers to the contact between LEU167 and LEU395. 
Panel C corresponds to the contact between the catalytically active residue and 
the glucose unit S$_{-1}$, where the hydrolysis reaction takes place. 
In all cases A, B and C, $\langle d_{ij} \rangle$ decreases with increasing $\kappa$. 
In contrast, panel D shows that the average distance between 
TRP205 and S$_{-2}$ increases with $\kappa$. 
For $\kappa$ in the range between 0 and 8~kcal/(mol~{\AA}$^{2}$), 
which we have studied here, the changes in the average distance, $\delta \langle d_{ij} \rangle$, 
are of the order of 0.1~{\AA}. 
These distance changes are clearly larger than the statistical errors of $\langle d_{ij} \rangle$, 
which we have estimated based on nine independent MD trajectories. 

\begin{figure}[h]
\begin{center}
\scalebox{0.58}{\includegraphics{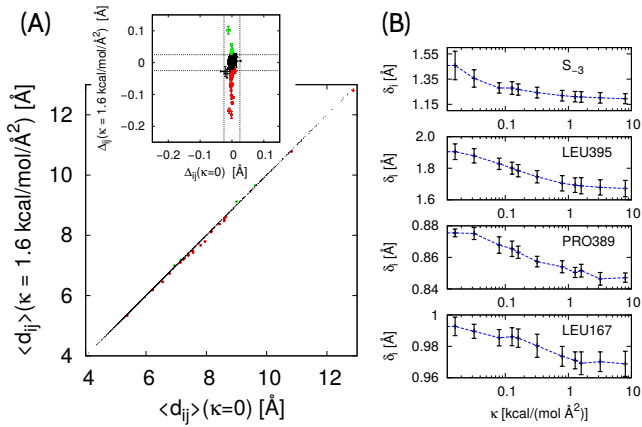}}
\caption{\label{dij_cross_plot} Results of the coarse-grained structure-based MD simulations. 
(A) $\langle d_{ij} \rangle$ computed for $\kappa = 0$, {\sl i.e.}, when only one spring 
with stiffness $\kappa_0$ is attached to residue S$_{-3}$, {\sl versus}  $\langle d_{ij} \rangle$ 
for $\kappa = 1.6$~kcal/mol/{\AA}$^2$. 
The elongated contacts are marked in red. The shortened contacts are marked in green. 
Inset: $\Delta_{ij} ( \kappa =0 )$ versus $\Delta_{ij} ( \kappa = 1.6$~kcal/mol/{\AA}$^2 )$, 
where $\Delta_{ij} ( \kappa ) = \langle d_{ij} \rangle (\kappa) - \langle d_{ij} \rangle_{\rm free}$ 
and $\langle d_{ij} \rangle_{\rm free}$ denotes the average inter-residue distances 
in a free CD which is not attached to any spring. 
(B) Root-mean-square fluctuations $\delta_i$ of the indicated residues 
as a function of the spring constant $\kappa$. 
These residues are depicted in Fig.~\ref{contacts_pic}. }
\end{center}
\end{figure}

In the limiting case of $\kappa = 0$, {\sl i.e.}, 
when the linker-imitating spring is not applied to the CD, 
$\langle d_{ij} \rangle$ are observed to have the same values 
-- within the statistical error -- 
as in the case of a free CD with $\kappa = 0$ and $\kappa_0 = 0$, {\sl i.e.}, 
when no springs are applied to the CD. 
This result can be seen in the inset of Fig.~\ref{dij_cross_plot}A, where 
$\Delta_{ij} ( \kappa ) = \langle d_{ij} \rangle ( \kappa ) - \langle d_{ij} \rangle_{\rm free}$ 
whereas $\langle d_{ij} \rangle_{\rm free}$ denote the average distances 
between residue pairs in the free CD.  For $\kappa > 0$, most of the native contacts  
are also practically unaffected by the attachment of the two springs. 
However, there is a group of about five native contacts that are observed to become longer 
as a result of applying the linker-imitating spring. 
They are marked in Fig.~\ref{dij_cross_plot}A in green. 
There is also a group of over twenty native contacts that become distinctly shorter upon 
applying this spring. These contacts are marked in red in Fig.~\ref{dij_cross_plot}A. 
The amino acids that form these contacts are shown as blue sticks in Fig.~\ref{contacts_pic}. 
It is interesting to note that the contact-length changes 
are observed to occur in many residue pairs in the vicinity of the substrate. 
This observation implies that variations in $\kappa$ 
lead to changes in the geometry of the substrate-binding region. 
We predict these geometrical changes to be rather small 
because $\delta \langle d_{ij} \rangle$ are of the order of 0.1~{\AA} only. 
However, in general, the catalytic activity of enzymes requires 
very precise positioning of substrate molecules in the active site. 
We therefore suggest that the linker-induced, small structural rearrangements of 
the active site cleft may influence the activity of Cel8A. 

We also performed MD simulations of the CD without the substrate molecule 
in the catalytic cleft. The PDB structure 1CEM was used as input for these simulations, 
and only one spring was attached to the CD, {\sl i.e.}, 
the one at the C-terminus with the effective stiffness $\kappa$. 
We find that regardless of the value of $\kappa$, 
the average distances $\langle d_{ij} \rangle$ are practically the same 
as in the case of the free catalytic domain that is not attached to any spring. 
These simulation results show that the linker-induced changes 
in the active site conformations can occur only in the bound state. 
We note that these conformational changes result from the restrains 
that are imposed on the mini-cellulosome by its attachment 
to cellulose chains at two distinct sites. 

We also analyzed root-mean-square fluctuations of individual residues, 
$\delta_i$, as defined by Eq.~(\ref{eq_rmsf}). 
The fluctuations of the great majority of residues 
are essentially unaffected by the stiffness of the second spring. 
Several exceptions are shown in Fig.~\ref{dij_cross_plot}B. 
We note that the influence of the linker-imitating spring on $\delta_i$ 
is local -- only the spring attachment sites (LEU395 and S$_{-3}$) and its 
nearest neighbors exhibit a dependence of $\delta_i$ on $\kappa$. 
Specifically, $\delta_i$ decreases with increasing $\kappa$ for these residues. 

As mentioned before, the structural changes in Cel8A are relatively small 
(variations in $\langle d_{ij} \rangle$ do not exceed 0.2~{\AA}; 
variations in $\delta_i$ are below 0.3~{\AA}) but very non-local 
(they propagate throughout the entire enzymatic domain). 
These small structural changes should be captured correctly by 
the structure-based model because its predictions are expected to be most 
accurate in a vicinity of the native state.

\subsubsection{Results of all-atom MD simulations \label{AA_results_WT}}

To validate the results of the coarse-grained MD simulations, 
we performed several all-atom MD simulations with explicit water, 
as described in Methods section \ref{methods_MD}. 
We used NAMD version 2.9 with CHARMM force fields and TIP3P water model. 
As in the coarse-grained simulations, one spring was attached to 
atom C4 of the glucose unit S$_{-3}$, and another one to 
to the $\alpha$-C atom of the amino-acid residue LEU395. 

\begin{figure}[h]
\begin{center}
\scalebox{0.72}{\includegraphics{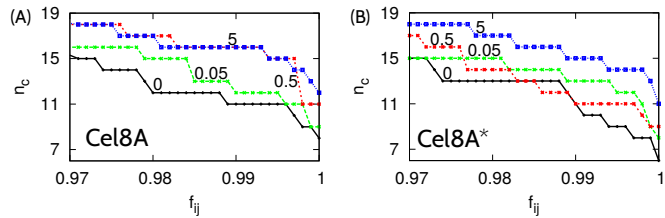}}
\caption{\label{nc_fij} 
The number of residue contacts between the endoglucanase and the cellopentaose, $n_{\rm c}$, 
which are present in a given fraction $f_{ij}$ of the recorded frames. 
These results correspond to the short, 2~ns, all-atom MD simulation trajectories. 
The labels 0, 0.05, 0.5 and 5 indicate the stiffness of the C-terminal spring, $\kappa$, 
given in the units of kcal/(mol {\AA}$^{2}$), which 
correspond to the black, green, red and blue lines, respectively. 
Panels A and B correspond to the wild type (Cel8A) and mutant (Cel8A*) enzyme, respectively. }
\end{center}
\end{figure}

To start, we performed four independent, short simulations, 2~ns each, 
in which the virtual spring attached to the C-terminus had the stiffness 
$\kappa$ of 0, 0.05, 0.5 and 5~kcal/(mol~{\AA}$^{2}$), respectively. 
The simulation structures were recorded every 1~ps, which 
resulted in generating 2000 'frames' for each of the four values of $\kappa$. 
Any 'frame' represented a simulation structure at a given time instant between 0 and 2~ns. 
Next, each of the frames was analyzed individually by an external program 
to determine the instantaneous contacts between the enzyme and its substrate. 
The contacts were identified using the overlap criterion. 
It turns out that some of the protein-sugar contacts persist throughout all of the frames. 
However, most of the contacts exist only in a fraction of the recorded frames. 
We denote this fraction by $f_{ij}$, where the subscript $ij$ labels the residues forming the contact. 
Fig.~\ref{nc_fij}A shows the number of contacts between Cel8A and its substrate, $n_{\rm c}$, 
as a function of $f_{ij}$ for $ \kappa $=0, 0.05, 0.5 and 5~kcal/(mol~{\AA}$^{2}$). 
For a given fraction $f_{ij}$, $n_{\rm c}$ generally increases with $\kappa$, 
especially for the most stable contacts with $f_{ij} > 0.95$. 
This result indicates that the stiffness of the linker-imitating spring affects 
the interactions between Cel8A and its substrate on the time scale of a few nanoseconds. 
Generally, increasing $\kappa$ leads to an increased stability of instantaneous contacts. 

We also generated three longer trajectories, 30~ns each, in which the stiffness of 
the linker-imitating spring, $\kappa$, was taken to have the following values: 
0 (meaning no spring attached to LEU395), 
0.25~kcal/(mol~{\AA}$^{2}$) (soft spring) and 
5~kcal/(mol~{\AA}$^{2}$) (stiff spring). 
Based on these longer trajectories, we computed the average distances 
between the $\alpha$-C and C4 atoms in the residues forming 
the native contacts between Cel8A and its substrate. 
Our goal was to directly compare these distances with the contact-length changes 
that we observed in the coarse-grained simulations. 
Examples of such a comparison are shown in Fig.~\ref{dij}C and D, which correspond to 
the native contacts GLN95-S$_{-1}$ and TRP205-S$_{-2}$, respectively. 
The results of the all-atom and coarse-grained simulations seem to be consistent. 
However, the statistical errors on $\langle d_{ij} \rangle$ 
from all-atom simulations are larger than the effect of contact-length changes. 
The significant statistical errors reflect most likely insufficient sampling 
in the all-atom simulations. In fact, the time scales of 
the all-atom (30~ns) and coarse-grained (0.5~ms) simulations are incomparable. 
Nevertheless, the all-atom MD simulations turn out to be very useful for studies 
on the effect of mutations in Cel8A, which we will discuss in the next section.

\subsection{Intra-domain contacts within the catalytic domain of Cel8A* \label{intra_domain_Cel8A*}}

We use FoldX \cite{foldX} to generate an atomic structure of the CD of 
the four-point mutant Cel8A*. In this approach, the four mutations 
(K276R, G283P, S329G, S375T) are assumed to have only a local effect on 
the structure of the CD. Specifically, the structures of 
the backbone chains of Cel8A and Cel8A* are assumed to be identical. 

\subsubsection{Results of coarse-grained structure-based simulations}

Using the overlap criterion \cite{Wolek} we determine native contacts 
within the mutant structure. The overlap criterion predicts 1117 contacts between 
amino acids and 30 contacts between the enzyme and its substrate.
The four mutations introduce nine new contacts and cause four contacts to vanish. 
However, they do not change the native contacts between the protein and the substrate. 

\begin{figure*}
\begin{center}
\scalebox{0.3}{\includegraphics{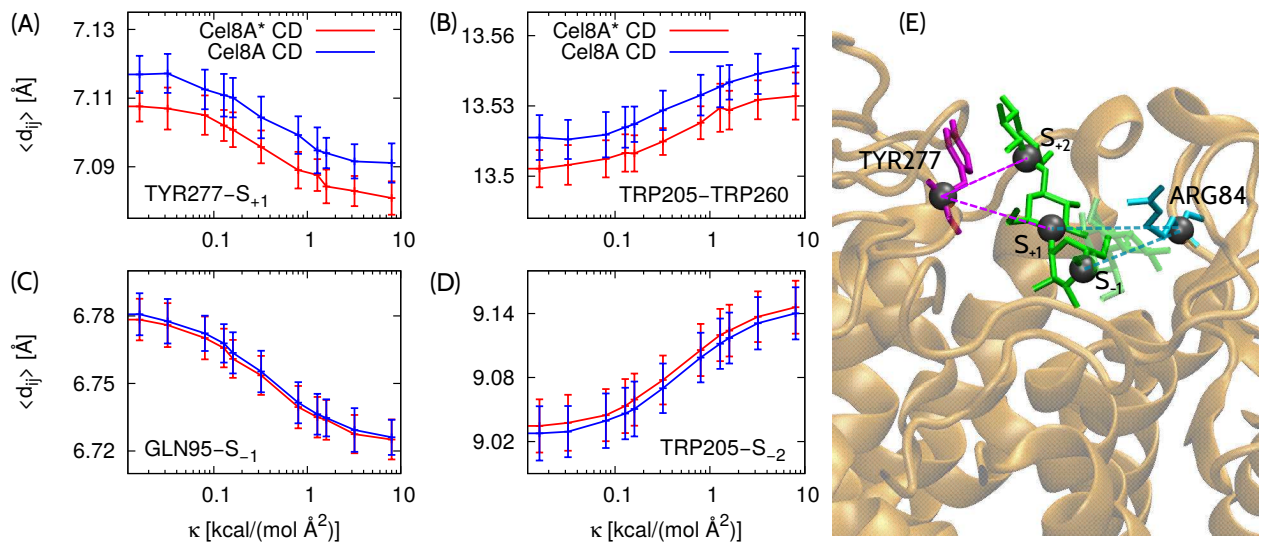}}
\caption{\label{diff} 
(A-D) $\langle d_{ij} \rangle $ as a function of $\kappa$ for indicated residue pairs 
within the CD of Cel8A (blue lines) and Cel8A* (red lines). 
These distances were determined in the coarse-grained MD simulations. 
Note that the logarithmic scale is used on the horizontal axes. 
(A) Simulation results corresponding to the contact between TYR277 and 
the sugar unit S$_{+1}$. This contact is marked as a dashed line in panel E. 
(B) Simulation data corresponding to the amino-acid contact between TRP205 and TRP260. 
The former residue participates in binding of the cellopentaose in the catalytic cleft. 
(C and D) The same residue pairs as in panels C and D of Fig.~\ref{dij}. 
These contacts are predicted by the coarse-grained MD simulations to be 
unaffected by the mutations. (E) Geometry of the catalytic cleft. 
Cellopentaose is shown in green, TYR277 in purple, and ARG84 in cyan. 
The gray spheres mark the $\alpha$-C atoms of the two amino acids 
and the C4 atoms of the sugar units S$_{-1}$, S$_{+1}$ and S$_{+2}$.  }
\end{center}
\end{figure*}

To study the equilibrium properties of the mutant enzyme (Cel8A* CD) 
in the vicinity of its native state, we used the same coarse-grained 
structure-based model as for the simulations of the wild-type enzyme (Cel8A CD). 
Two virtual spring were attached to Cel8A* CD in complex with cellopentaose: 
the first one to unit S$_{-3}$ of the substrate molecule 
and the second one to the C-terminal residue (LEU395). 
The attachment sites for the springs were thus the same as those in Cel8A CD. 

We performed coarse-grained MD simulations of this system for several different 
values of $\kappa$ ranging from 0 to 8~kcal/(mol~{\AA}$^{2}$). 
We analyzed the average distances between the residue pairs forming native contacts, 
$\langle d_{ij} \rangle$, and found that the great majority of them 
were unaffected by the four mutations. 
However, we notice several exceptions, two of which are depicted in Fig.~\ref{diff}. 
The blue and red lines correspond to Cel8A and Cel8A*, respectively. 
Panel A shows $\langle d_{ij} \rangle$ as a function of $\kappa$ for 
the native contact between residue TYR277 and the glucose unit S$_{+1}$. 
Panel B corresponds to $\langle d_{ij} \rangle ( \kappa )$ 
for the residue pair TRP205 and TRP260. 
The former amino-acid residue participates in the binding of 
the cellopentaose in the catalytic cleft. 
In both cases A and B, systematic differences in 
$\langle d_{ij} \rangle ( \kappa )$ are observed between Cel8A CD and Cel8A* CD. 
In contrast, panels C and D -- which correspond to the same residue pairs as 
panels C and D in Fig.~\ref{dij} -- show two examples where 
$\langle d_{ij} \rangle ( \kappa )$ are practically identical for Cel8A CD and Cel8A* CD. 
These results indicate that the stiffness of the linker-imitating spring can 
affect the geometry of the catalytic cleft region in Cel8A and Cel8A* in different ways. 

The contacts between the enzyme and its substrate are the same in Cel8A* CD 
and Cel8A CD. However, the average lengths of several of these contacts are different. 
Namely, the contact between TYR277 and S$_{+1}$ is found to shorten upon introducing the mutations, 
as shown in Fig.~\ref{diff}A. There are also three contacts that are observed to become 
slightly longer as a result of the mutations: ARG84-S$_{+1}$, ARG84-S$_{-1}$ and ALA150-S$_{-1}$. 
As shown in Fig.~\ref{diff}E, TYR277 and ARG84 are located on the opposite sides of 
the catalytic cleft. Therefore, the coarse-grained simulations indicate that 
the four mutations lead to a small shift of 
units S$_{+1}$ and S$_{-1}$ from one side of the catalytic cleft to another. 
We note, however, 
that our assumption about identical structures of the backbone chains of Cel8A and Cel8A* 
may not be correct. In fact, the four mutations may lead to non-local changes in 
the CD structure. In such cases, there would be more differences 
in the contact maps for Cel8A and Cel8A*. We would then expect the differences 
in $\langle d_{ij} \rangle ( \kappa )$ between Cel8A and Cel8A* to be even more evident.

\subsubsection{Results of all-atom MD simulations \label{AA_results_mut}}

To test the predictions of the coarse-grained model, 
we performed all-atom MD simulations with explicit water 
as described in Methods section \ref{methods_MD}. 
We performed four short simulations, 2~ns each, 
and analyzed the instantaneous contacts 
using the same method as in the case of Cel8A simulations. 
Fig.~\ref{nc_fij}B shows the number of contacts between Cel8A* and its substrate, 
$n_{\rm c}$, which are observed in a given fraction $f_{ij}$ of the recorded frames. 
The black, green, red and blue lines correspond to 
$\kappa $=0, 0.05, 0.5 and 5~kcal/(mol~{\AA}$^{2}$), respectively. 
As in the case of the wild type enzyme, $n_{\rm c}$ as a function of $f_{ij}$ 
generally increases with increasing $\kappa$. 
This result indicates that increasing the stiffness of the linker-imitating spring 
leads to an increased stability of the instantaneous contacts 
on the time scales of a few nanoseconds. 

\begin{figure*}
\begin{center}
\scalebox{0.68}{\includegraphics{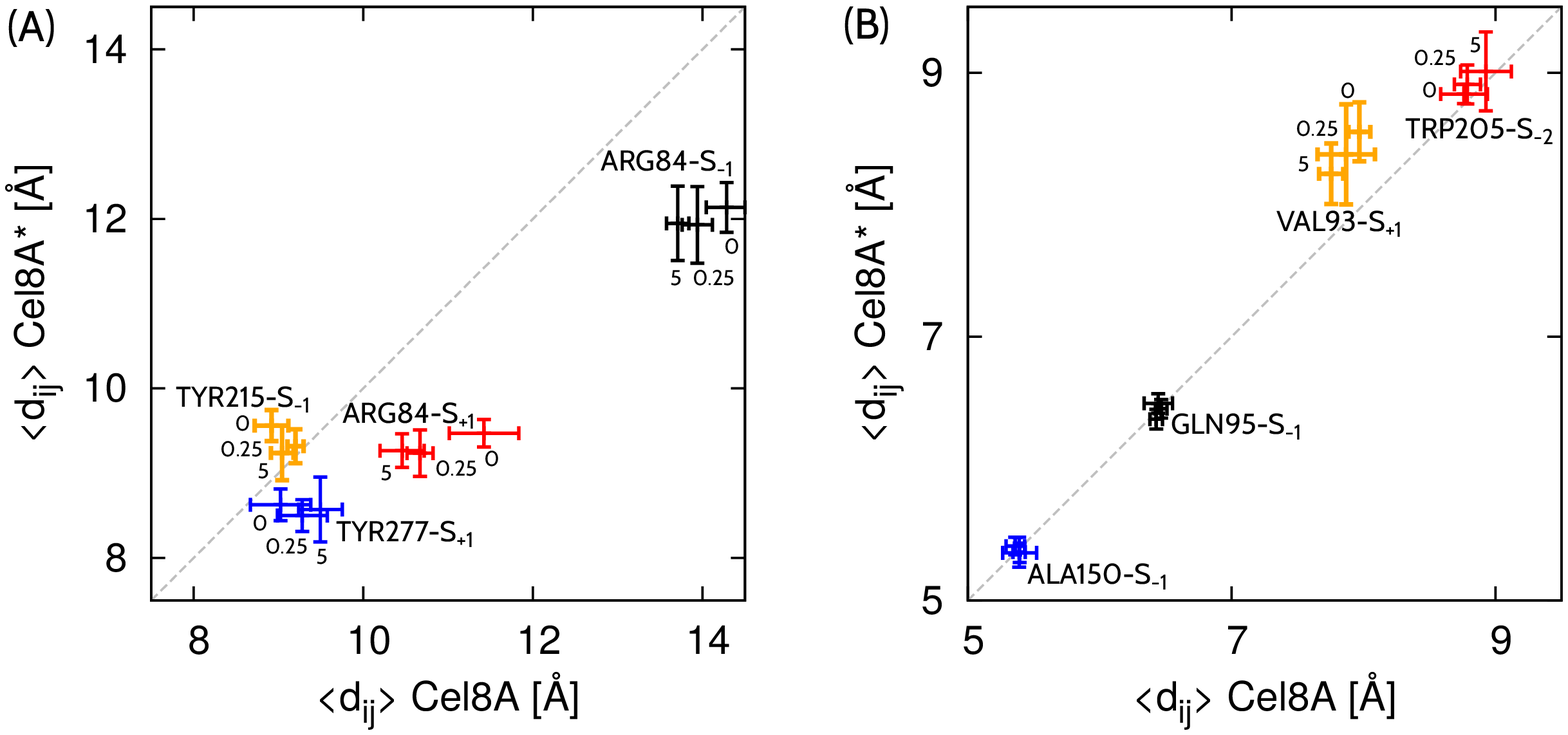}}
\caption{\label{cross_WT_mut} 
$\langle d_{ij} \rangle$ in Cel8A versus $\langle d_{ij} \rangle$ in Cel8A* for 
indicated contacts between the endoglucanase and the cellopentaose molecule. 
These distances were measured in the 30~ns MD simulations. 
The labels 0, 0.25 and 5 indicate the values of $\kappa$ 
given in the units of kcal/(mol~{\AA}$^{2}$). 
(A) Four contacts for which $\langle d_{ij} \rangle ( \kappa )$ 
in Cel8A and Cel8A* are clearly different. 
Three of the indicated contacts (TYR277-S$_{+1}$, ARG84-S$_{+1}$ and ARG84-S$_{-1}$, 
see Fig.~\ref{diff}D) have been predicted by the course-grained simulations 
to have different lengths in Cel8A and Cel8A*. 
For example, the blue data points indicate the contact between TYR277 and S$_{+1}$, 
which corresponds to Fig.~\ref{diff}A. 
(B) Four contacts for which $\langle d_{ij} \rangle ( \kappa )$ are practically unaffected, 
or affected only weakly (VAL93-S$_{+1}$), by the mutations in the endoglucanase. 
The black and red data points indicate GLN95-S$_{-1}$ and TRP215-S$_{-2}$, 
respectively, which correspond to Fig.~\ref{diff}C and D.   }
\end{center}
\end{figure*}

As discussed in the previous subsection, the course-grained simulations 
predict that four contacts between the enzyme and its substrate 
change their lengths when Cel8A is mutated to Cel8A*: 
TYR277-S$_{+1}$, ARG84-S$_{+1}$, ARG84-S$_{-1}$, and ALA150-S$_{-1}$. 
We performed 30~ns all-atom MD simulations of Cel8A CD and Cel8A* CD, 
with $\kappa$=0, 0.25 and 5~kcal/(mol~{\AA}$^{2}$), 
and measured the average distances between the $\alpha$-C and C4 atoms in 
these four pairs of residues. 
We found that the average distances were indeed different in Cel8A and Cel8A*, 
as shown in Fig.~\ref{cross_WT_mut}. 
Both the coarse-grained and all-atom simulations show that 
TYR277-S$_{+1}$ shortens upon introducing the mutations. 
However, the magnitude of the length changes is larger in the all-atom simulations, 
see Fig.~\ref{diff}A and the blue data points in Fig.~\ref{cross_WT_mut}A. 
Moreover, the coarse-grained simulations predict that ARG84-S$_{+1}$ and ARG84-S$_{-1}$ 
become slightly longer as a result of the mutations, which suggests 
a small shift of units S$_{+1}$ and S$_{-1}$ from one side of the catalytic cleft to another.
This prediction is inconsistent with the results of the all-atom simulations, 
in which ARG84-S$_{+1}$ and ARG84-S$_{-1}$ become significantly shorter, by about about 2~{\AA}, 
upon introducing the mutations into the catalytic domain. 
Visual inspection of the all-atom trajectories reveals that the rings S$_{-1}$, S$_{+1}$ and 
S$_{+2}$ are embedded somewhat deeper in the catalytic cleft of Cel8A* in comparison to 
the location of the cellopentaose molecule in the catalytic cleft of Cel8A. 

The all-atom simulations show that there are many contacts between the enzyme and its substrate 
which -- within the simulation error -- do not change their lengths when Cel8A is mutated to Cel8A*.  
Three examples are shown in Fig.~\ref{cross_WT_mut}B, where the black and red data points 
correspond to GLN95-S$_{-1}$ and TRP215-S$_{-2}$, respectively. 
These two contacts were found to be unaffected by the mutations also in the coarse-grained simulations. 
In contrast, the contact between ALA150 and S$_{-1}$ seems to be unchanged in the all-atom simulations 
(blue data points in Fig.~\ref{cross_WT_mut}B) but it is found to elongate by about 0.02~{\AA} 
in the coarse-grained simulations. However, such small changes in the contact length can not 
be identified in the all-atom simulations due to the statistical errors and insufficient sampling. 

Despite the fact that the coarse-grained and all-atom simulations give consistent predictions 
about the length changes of certain native contacts, they do not agree qualitatively in general. 
The major reason for this disagreement is most likely the assumption underlying the coarse-grained model 
that the backbone chains of Cel8A and Cel8A* are identical. 
Nevertheless, both the coarse-grained simulations and the all-atom simulations show that 
the four mutations in the endoglucanase as well as variations in the stiffness of 
the linker-mimicking spring, $\kappa$, lead to noticeable changes in 
the average lengths of certain contacts between the enzyme and its substrate. 
Moreover, the contact-length changes due to variations in $\kappa$ 
can be different in Cel8A and Cel8A*, see Figs.~\ref{diff} and \ref{cross_WT_mut}.

\subsection{Estimates on the stiffness of disordered linkers \label{linker_stiffness} }

In our recent work on conformational variability of the multi-domain 
xylanaze Z (XynZ) of {\sl C. thermocellum}, we determined the average 
end-to-end distance, $\langle d_{\rm ee} \rangle$, and the root-mean-square 
end-to-end distance, $\langle d_{\rm ee}^2 \rangle^{1/2}$ for the three disordered 
linkers of the sequential length of 6, 12 and 24 amino-acid residues \cite{XynZ2015}. 
From the variance of the end-to-end distance, 
$\langle d_{\rm ee}^2 \rangle - \langle d_{\rm ee} \rangle^2$,
we can estimate the effective spring constant, $\kappa$, which corresponds to such distance 
variations as induced by thermal fluctuations. In fact, equipartition of energy 
implies that 
\begin{equation}
\frac{1}{2} \kappa 
\left( \langle d_{\rm ee}^2 \rangle - \langle d_{\rm ee} \rangle^2
\right) = \frac{3}{2} k_B T
\end{equation}
and thus
\begin{equation}
\kappa = \frac{3 k_B T}{\langle d_{\rm ee}^2 \rangle - \langle d_{\rm ee} \rangle^2}
\end{equation}
At room temperature, $T=300$~K, we obtain $ \kappa $=0.26~kcal/(mol~{\AA}$^2$), 
$ \kappa $=0.07~kcal/(mol~{\AA}$^2$) and $ \kappa $=0.02~kcal/(mol~{\AA}$^2$) 
for the three linkers that are formed of 6, 12 and 24 amino-acid residues, respectively. 
Note that $\kappa$ exhibits a 14-fold decrease when 
the sequential length of the linkers increases from 6 to 24. 
It is because of this effect that we use the logarithmic scale in Figs.~\ref{dij} and \ref{diff}. 

These values of $\kappa$ are obtained for the three linkers within XynZ that have a specific 
amino-acid sequence. In general, $\langle d_{\rm ee}^2 \rangle - \langle d_{\rm ee} \rangle^2$ 
should depend not only on the sequential length of the linkers but also on their 
composition. Therefore, amino-acid mutations within linkers can lead to changes in 
the effective stiffness $\kappa$.

\section{Methods}

In this work we have employed three types of simulations: 
(i) coarse-grained Monte Carlo simulations that are used to characterize 
the conformational ensemble of the full-length Cel8A-ScafT complex; 
(ii) coarse-grained, structure-based Molecular Dynamics simulations 
that are used to investigate the influence of the linker-imitating spring on 
the structure of the catalytic domain of Cel8A and Cel8A*; and 
(iii) all-atom Molecular Dynamics simulations that give insights into 
the effect of the four-point mutation on the enzyme-substrate interactions. 
These simulation methods will we described in the following three subsections. 

\subsection{Coarse-grained simulations of Cel8A-ScafT and Cel8A*-ScafT \label{methods_CG} }

To sample physical conformations of the full-length Cel8A-ScafT complex, 
we use a coarse-grained model equipped with a transferable energy function 
that has been developed to simulate protein binding \cite{Hummer2008}. 
In the framework of this model, amino acid residues are represented as spherical
beads centered at the $\alpha$-C atoms.  
The interactions between the residue beads are described by short-range 
Lennard-Jones (LJ) type potentials and long-range electrostatic potentials. 
The electrostatic interactions are described by a Debye-H\"uckel-type potential. 
The LJ-type potentials are characterized by (i) amino-acid dependent interaction strengths, 
$\tilde{\epsilon}_{ij}$, which are adapted from knowledge-based statistical contact potentials 
introduced by Miyazawa and Jernigan \cite{MJ1996}, 
and (ii) residue-dependent interaction radii 
$\tilde{\sigma}_{ij} = ( \tilde{\sigma}_i + \tilde{\sigma}_j ) / 2$, 
where $\tilde{\sigma}_i$ is the van der Waals diameter of residue $i$, which is calculated 
from the van der Waals volume by assuming a spherical shape for the residue, 
see Table~5 in Ref.~\cite{Hummer2008}. 

Folded protein domains are represented as rigid bodies whereas 
flexible linker peptides connecting the domains are represented as 
polymers of amino acid beads with bending, stretching and torsion potentials.
A detailed description of the model can be found in Ref.~\cite{Hummer2008}. 
The full-length model of the Cel8A-ScafT complex comprises three rigid domains: 
(i) the catalytic domain of endoglucanase Cel8A (PDB code 1CEM), 
(ii) the carbohydrate-binding module of family 3a (CBM3A with the PDB code 4JO5), 
and (iii) the dockerin domain of the endoglucanase Cel8A in complex with 
the cohesin-3 domain of the scaffoldin CipA. To the best of our knowledge, there is no PDB entry with 
the structure of the dockerin-cohesin complex that corresponds to the rigid body (iii). 
For this reason we use a homology model generated by SWISS-MODEL \cite{Biasini2014} 
with the PDB structures 1AOH and 4DH2 as templates for the cohesin (94 percent sequence identity) 
and dockerin (43 percent sequence identity), respectively. In this homology model, 
the dockerin-cohesin interface is analogous to that found in the PDB structure 4DH2. 
In addition to the three rigid bodies, the full-length model of Cel8A-ScafT comprises 
two flexible linkers, as shown in Fig.~\ref{complex_struc}. 
The sequential length of the Cel8A and ScafT linkers is 17 and 25 amino acids, respectively. 
Their amino acid sequences are given in Supplementary Information. 
We generated the initial configurations of the disordered linkers using ModLoop \cite{modloop}. 

We performed Replica Exchange Monte Carlo (REMC) simulations 
using in-house software with replicas at sixteen different temperatures $T_i$ 
given by $T / T_i = 0.6, 0.64, \ldots, 1.16, 1.2$ relative to the room temperature $T$=300~K. 
The basic MC steps were rigid body translational and rotational moves on each domain. 
For flexible linker peptides, in addition to local MC moves on each residue, 
crank shaft moves were employed to enhance sampling. 
The probabilities of the MC moves were governed by the Metropolis criterion. 
After the initial $10^6$ MC sweeps for equilibration, 
$10^7$ MC sweeps were performed for data acquisition. 
The protein conformations were saved every $10^3$ MC sweeps, 
which gave us an ensemble of $10^4$ conformations for further analysis. 
In order to check that the simulation results were not biased by the initial 
conformation, we performed five independent simulation runs 
that were started with different seeds for the random number generator. 

\subsection{Coarse-grained simulations of the catalytic domain \label{methods_Go} }

To sample near-native conformations of the catalytic domain of endoglucanase Cel8A
in complex with its substrate (cellopentaose), we use our coarse-grained 
structure-based model \cite{Hoang2,SikoraPLOS2009,Proteins2014,Poma2015}, 
in which amino-acid residues and sugar units are represented 
by single beads centered on their $\alpha$-C and C4 atoms, respectively. 
The beads are tethered together into two chains by strong harmonic potentials 
with the spring constant of $100 \, \epsilon /$\AA$^{2}$, where $\epsilon$ 
is the depth of the potential well associated with the native
contacts, which serves as the basic energy scale in this model. 
The chain of amino-acid beads corresponds to the enzyme and 
the chain of glucose-ring beads corresponds to its substrate. 
The native contacts are identified using an overlap criterion \cite{Wolek} 
applied to the coordinates of all heavy atoms in the native structure (PDB code 1KWF \cite{1KWF}). 
The van der Waals radii for the heavy atoms of amino acids and sugar residues 
are taken from Refs.~\cite{Tsai} and \cite{Poma2015}, respectively. 
The amino acid pairs that are very close sequentially, $(i,i+1)$ and $(i,i+2)$, 
are excluded from the contact map \cite{SikoraPLOS2009}. 
The contacts between the protein and the substrate are treated 
in the same manner as the contacts within the protein 
as both sets are dominated by hydrogen bonds \cite{Proteins2014}. 

The interactions within the native contacts are described by the Lennard-Jones potential 
\begin{equation}
V^{\rm NAT} (r_{ij}) = 4 \epsilon \left[ \left( \frac{\sigma_{ij}}{r_{ij}} \right)^{12} 
- \left( \frac{\sigma_{ij}}{r_{ij}} \right)^{6}  \right]
\label{eq:6-12}
\end{equation}
Here, $r_{ij}$ is the distance between residue beads $i$ and $j$ forming the native contact, 
and the parameters $\sigma_{ij}$ are chosen so that each contact in the native structure is 
stabilized at the minimum of the Lennard-Jones potential. 
The value of $\epsilon$ is approximately given by $(1.6 \pm 0.4)$ kcal/mol, 
as has been determined by comparing simulational results to the experimental
ones on a set of 38 proteins \cite{SikoraPLOS2009}. 
The interactions between the pairs of residue beads 
that do not form native contacts are purely repulsive 
and given by a truncated and shifted Lennard-Jones type potential 
\begin{equation}
V^{\rm NON-NAT} (r_{ij}) = 4 \epsilon \left[ \left( \frac{\sigma_{0}}{r_{ij}} \right)^{12} 
- \left( \frac{\sigma_{0}}{r_{ij}} \right)^{6} + \frac{1}{4} \right] 
\quad {\rm for} \quad r_{ij} \le r_0 
\end{equation}
and $V^{\rm NON-NAT} (r_{ij}) = 0$ for $r_{ij} > r_0$.  
Here, $\sigma_{0} =  r_0 / \sqrt[6]{2}$ and $r_0 = 4$~{\AA}. 
The potential energy comprises also harmonic terms that favor
the native values of the local chirality \cite{Kwiecinska}. Specifically, we use 
\begin{equation}
V^{\rm C} = \frac{k_c}{2} \sum_{i=2}^{N-2} \left( C_i - C_i^{\rm NAT} \right)^2 
\quad {\rm with} \quad 
C_i = \frac{\left( \vec{w}_{i-1} \times \vec{w}_{i} \right) \cdot \vec{w}_{i+1} }{ d_{i}^{3} }
\end{equation}
Here, $C_i^{\rm NAT}$ and $C_i$ denote the chirality of residue $i$ in 
the native and instantaneous conformation, respectively, 
$\vec{w}_i = \vec{r}_{i+1} - \vec{r}_{i}$, and $d_i = | \vec{w}_i |$ 
is the distance between subsequent residue beads $i$ and $i+1$. 
We take $k_c = \epsilon$, as in Ref.~\cite{Kwiecinska}. 
By construction, the global minimum of the potential energy 
corresponds to the native structure. 

The solvent is implicit and the system evolves in time according 
to the Langevin dynamics. The overall force acting on a particular bead $i$ 
is a sum of three terms: (i) the direct force $\vec{F}_i$ that derives from all 
the potential terms, (ii) the damping force that is proportional to the velocity of 
the bead, and (iii) the random force, $\vec{\Gamma}_i$, that represents thermal noise. 
The corresponding equations of motion 
\begin{equation}
m \frac{ {\rm d}^2 \vec{r}_i }{ {\rm d} t^2} = 
\vec{F}_i - \gamma \frac{ {\rm d} \vec{r}_i }{ {\rm d} t} + \vec{\Gamma}_i
\end{equation}
are solved by the fifth order predictor-corrector algorithm with 
the time step of $0.005 \, \tau$ using in-house software. 
Here, $\gamma$ is the damping coefficient and all beads are assumed to have the same mass $m$. 
The dispersion of the noise is given by 
$\sqrt{ 2 \gamma k_B T \,}$, where $k_B$ is the Boltzmann constant 
and $T$ denotes the temperature. 
All of the simulations reported here are performed at $k_B T = 0.3 \, \epsilon$, 
which is near-optimal in folding kinetics and is of order of the room temperature.
The damping coefficient is set to $\gamma = 2 m / \tau$. This value corresponds
to the overdamped case -- practically Brownian dynamics -- and
the characteristic time scale, $\tau$,
is of the order of 1~ns, as argued in Refs.~\cite{Thirumalai,flow}. 
The average distances $\langle d_{ij} \rangle$ between pairs of beads $(i,j)$ 
within native contacts are computed from nine independent trajectories 
of $6 \times 10^4 \, \tau$ each. The trajectories start from the native state 
with different seeds for the random number generator, and the first $10^4 \, \tau$ 
are used for equilibration and excluded from computing $\langle d_{ij} \rangle$. 

To quantify the local flexibility of the enzyme, we compute the root-mean-square 
fluctuation (RMSF) of each residue 
\begin{equation}
\delta_i = \left[ \langle \vec{r}_i^{\; 2} \rangle -
\langle \vec{r}_i \rangle^2 \right]^{1/2}
\label{eq_rmsf}
\end{equation}
Here, $\vec{r}_i$ is the position of the $i$-th residue and the angle brackets denote 
the average after superimposing on the native structure. 

\subsection{All-atom simulations of the catalytic domain \label{methods_MD} }

The atomic coordinates of the catalytic domain in complex with the cellopentaose 
were taken from the crystal structure reported in Ref.~\cite{1KWF} (PDB code 1KWF). 
The atomic structure of the four-point mutant was generated using FoldX \cite{foldX}. 
The initial systems for all-atom MD simulations were prepared using VMD \cite{VMD}. 
Namely, the structures were solvated using Solvate Plugin version 1.5. 
Sodium and chloride ions were added to neutralize the simulated system 
and to reach a physiological ion concentration of 150~mM. 

The all-atom MD simulations were performed using NAMD 2.9 \cite{NAMD}. 
CHARMM22 force field \cite{CHARMM22} with CMAP correction \cite{CMAP} was used for proteins. 
The parameters of CHARMM36 force filed \cite{CHARMM36_sugar_1,CHARMM36_sugar_2,CHARMM36_sugar_3} 
were employed for the cellopentaose. 
The TIP3P water model was used \cite{TIP3P}. The simulations were carried out in the NPT ensemble. 
Temperature was kept at 300~K through a Langevin thermostat with a damping coefficient of 1/ps. 
Pressure was maintained at 1~atm using the Langevin piston Nose-Hoover method 
with a damping timescale of 50~fs and an oscillation timescale of 100~fs. 
Short-range non-bonded interactions were cut off smoothly between 10 and 12~{\AA}. 
Long-range electrostatic interactions were computed using the particle-mash Ewald method 
with a grid spacing of 1~{\AA}. Simulations were performed with an integration time step of 2~fs. 

Prior to the actual simulations, the system was relaxed in three subsequent steps: 
(i) relocation of atoms and water molecules with 10000 iterations of a conjugate gradient energy minimization; 
(ii) 0.1~ns simulation at temperature $T = 100$~K; and (iii) 0.1~ns simulation at $T = 200$~K. 
We performed several short, 2~ns simulations as well as longer, 30~ns simulations of 
both the wild type CD and the four-point mutant CD. 
The stiffness of the virtual spring attached to the 
$\alpha$-C atom of the C-terminal residue LEU395 
was taken to have several different values in these simulations,
as specified in Results sections \ref{AA_results_WT} and \ref{AA_results_mut}.

\section{Summary}

Significant efforts have been made to identify and implement mutations that 
improve stability and activity of individual domains of cellulosomes. 
However, the recent discoveries made by the Bayer group show that enhancing 
thermostability and activity of separate enzymatic domains need not 
necessarily lead to an improvement in overall degradation of cellulose by 
designer cellulosomes \cite{Bayer_CR_2014}. 
Our results indicate that mutations that change the effective stiffness of 
the disordered linkers can modulate the enzyme structure in a non-local manner. 
These mutations may involve insertions and deletions, which change the linker length, 
or amino-acid replacements that influence the persistence length of linkers. 
The resulting modulations of the enzyme structure may in turn 
imply changes in the activity of cellulosomal enzymes. 
Our results may thus help explain the reasons for the astonishing variability 
in sequences among the cellulosomal linkers. 
Our results may also shed light on the discovery made by the Bayer group 
that designer cellulosomes assembled with the native long-linker scaffoldins 
achieve higher levels of activity compared to those assembled 
with short-linker scaffoldins \cite{Bayer2013}. 

Disordered and flexible linkers are found in various multi-domain proteins \cite{Nussinov2016,JPCM2014}. 
It would be thus interesting to study whether the elastic properties of linkers 
could influence the activity of other multi-domain enzymes. 
Interestingly, in several regulatory proteins, linkers have been reported to serve 
not only as passive tethers but also as active modulators of 
the function elicited by adjacent domains \cite{BJlinkers2015,TsaiStructure2011,MelaciniPNAS2013}. 
In particular, recent experiments on protein kinase C show that inter-domain linkers 
play critical roles in stabilizing certain conformation \cite{Tom2015}.

\section*{Acknowledgements}

\noindent This research has been supported by 
the European Framework Programme VII NMP grant 604530-2 (CellulosomePlus), 
the ERA-NET-IB/06/2013 Grant FiberFuel funded by 
the National Centre for Research and Development in Poland, 
and the Grant 2014/15/B/ST3/01905 from the National Science Centre.
It was also co-financed by the Polish Ministry of Science and 
Higher Education from the resources granted for the
years 2014-2017 in support of international scientific projects.




\end{document}